# On-chip fully reconfigurable Artificial Neural Network in 16 nm FinFET for Positron Emission Tomography

Andrada Muntean, *Student Member,* Yonatan Shoshan, Slava Yuzhaninov, Emanuele Ripiccini, Claudio Bruschini*, Senior Member*, IEEE, Alexander Fish, Edoardo Charbon, *Fellow, IEEE*

*Abstract—* Smarty is a fully-reconfigurable on-chip feed-forward artificial neural network (ANN) with ten integrated time-to-digital converters (TDCs) designed in a 16 nm FinFET CMOS technology node. The integration of TDCs together with an ANN aims to reduce system complexity and minimize data throughput requirements in positron emission tomography (PET) applications. The TDCs have an average LSB of 53.5 ps. The ANN is fully reconfigurable, the user being able to change its topology as desired within a set of constraints. The chip can execute 363 MOPS with a maximum power consumption of 1.9 mW, for an efficiency of 190 GOPS/W. The system performance was tested in a coincidence measurement setup interfacing Smarty with two groups of five 4 mm × 4 mm analog silicon photomultipliers (A-SiPMs) used as inputs for the TDCs. The ANN successfully distinguished between six different positions of a radioactive source placed between the two photodetector arrays by solely using the TDC timestamps.

## I. INTRODUCTION

For a long time, artificial intelligence (AI) has been applied in different areas such as biology, economics, medical healthcare, automotive, etc. [1], [2], [3], [4]. AI makes use of different tools depending on the problem that needs to be solved, such as deep learning, ANN, probabilistic methods and many others [5], [6], [7]. ANNs, in particular, were initially inspired by the structure of the human brain. Neurons are at the core of our nervous system, whose connections are established through synapses. The connection between neurons is essential for learning, as it serves as a way in which information is sent between them. ANNs are based on the same approach, each of them featuring a set of neurons organized by layers and connections. Each connection has assigned a specific weight, whose role is to decide how much influence the input has on the output value. A typical feed-forward artificial neural network consists of several layers which are usually classified as input, hidden and output layers. Usually, the input layer receives the information coming from outside, in our case, from the TDCs, while the output layer provides the final result. However, there are many ways in which ANNs can be implemented. In the past, ANNs have found applicability in the medical field by assisting image reconstruction algorithms in enhancing image quality [8], [9], [10], [11], [12]. Considering that clinical diagnostic systems, such as magnetic resonance imaging (MRI), computed tomography (CT), or PET, need to handle large amounts of data that are usually processed offline, ANNs proved to be a valuable tool for this task. PET is a nuclear imaging technique heavily used in oncology for diagnostics, treatment, and monitoring of cancerous tumors. This technique uses radioisotopes, which are injected in the patient's body and, in some cases, concentrate in the area where the tumor is located. The radioisotope undergoes positron decay, and the emitted positron travels a short distance in the tissue until it interacts with an electron. This process, called annihilation, results in two annihilation photons emitted with a photon energy of 511 keV at 180 degrees forming a line-of-response (LOR). The emitted gamma rays are absorbed by scintillators and converted into visible photons which are then detected by optical photodetectors, as depicted in Fig. 1. The role of the PET

A. Muntean, the corresponding author, is with EPFL, Neuchâtel, NE 2002 Switzerland (e-mail: andrada.muntean@epfl.ch).
Y. Shoshan is with Bar-Ilan, Israel, Ramat Gan 5290002 Israel (e-mail: yonatan.shoshan@biu.ac.il).
S. Yuzhaninov is with Bar-Ilan, Israel, Ramat Gan 5290002 Israel (e-mail: Slava.Yuzhaninov@biu.ac.il).
E. Ripiccini is with EPFL, Neuchâtel, NE 2002 Switzerland (e-mail: emanuele.ripiccini@epfl.ch).
C. Bruschini is with EPFL, Neuchâtel, NE 2002 Switzerland (e-mail: claudio.bruschini@epfl.ch).
A. Fish is with Bar-Ilan, Israel, Ramat Gan 5290002 Israel (e-mail: alexander.fish@gmail.com).
E. Charbon is with EPFL, Neuchâtel, NE 2002 Switzerland (e-mail: edoardo.charbon@epfl.ch).

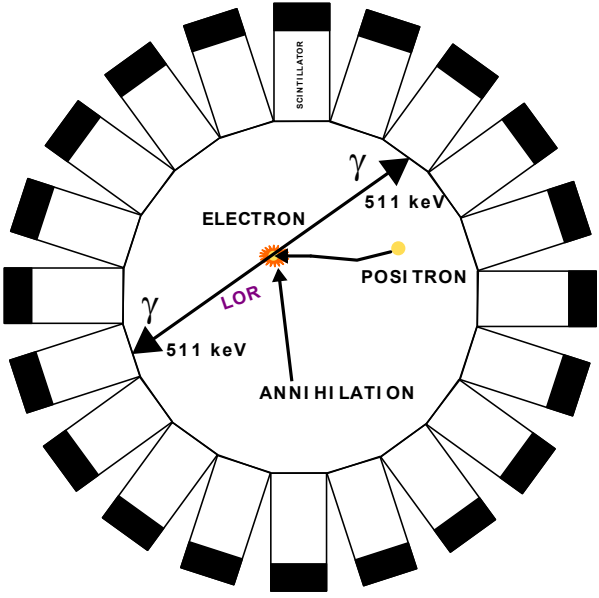

Fig. 1. Conceptual representation of a PET ring. The interaction between positron and electron (annihilation) results in two gamma rays emitted with a photon energy of 511 keV at 180 degrees along a LOR. The gamma rays reach first the scintillator, which converts them into visible photons. The visible photons are further detected by the photodetectors.

scanner is to acquire a subset of LORs and to reconstruct the most likely annihilation point in 3D. Time-of-flight PET (ToF-PET) systems, in particular, make use of timing information obtained from photodetectors to reconstruct the annihilation point with more advanced reconstruction algorithms, enabling enhanced image quality. Over the years, different techniques have been used to acquire the ToF-PET information (timing, energy to discriminate the 511 keV), such as constant fraction discrimination, leading edge discrimination or different estimation algorithms based on statistical models [13], [14]. More recently, the research focus shifted towards ToF estimation with ANNs. The work carried out in [13] presents a nine-layer off-chip convolutional neural network, which uses digitized waveforms as an input through constant-fraction discriminators to estimate ToF information. The ANN is trained in Matlab with millions of coincidence events acquired with a $^{22}$Na point source at different timing delays. In the end, an average timing precision of 32 ps was obtained by using this approach.

Another interesting study has been presented in [15]. The authors analyze the capabilities of monolithic scintillators with respect to timing and spatial resolution by using an off-chip neural network. The proposed ANN is implemented in FPGA and comprises thousands of neurons and coefficients while the readout electronics is composed of custom developed application specific integrated circuits (ASICs).

One of the main issues in PET is the large amount of data that is generated from thousands of photodetectors, and which needs to be transferred for external processing [16], [17], [18]. Moreover, the data has to be filtered to remove random and scattered events in order to enhance the final image quality.

In this paper, we propose an on-chip fully-reconfigurable ANN with the goal of reducing complexity and minimizing data throughput. Smarty comprises both timestamping circuitry and an ANN. Ten independent channels are directly connected to Smarty's reconfigurable ANN, which has been trained for the reconstruction of the position of a $^{22}$Na radioactive source placed in-between two photodetector arrays. The data from the ten input channels is reduced to a single word per frame and special cases, when not enough channels fired, are automatically discarded by the ANN. The topology of the ANN can be changed within certain design limits, i.e. a maximum of 1024 weights and biases and 128 neurons. Smarty proposes a fully-integrated, reconfigurable ANN used to reconstruct the position of a $^{22}$Na source along the X axis with floating point and quantized representation results presented in the end.

This work is organized as follows: in section II the on-chip neural modelling, which is the basis of the entire design, is presented. This is followed by the in-depth description of the Smarty system architecture in section III, where the TDCs and ANN implementation are discussed. Section IV presents the performance characterization results, where the TDCs, the genetic algorithm training, and the first ANN on-chip performance are validated. Section V presents the source position reconstruction measurements performed with Smarty's ANN, followed by section VI which provides the conclusions of this work.

## II. ON-CHIP NEURAL NETWORK MODELLING

The neural network was firstly described through a mathematical model before being implemented on-chip. The ANN's reconfigurability is encoded by means of a topology file, which contains information on the ANN's structure such as number of neurons, the number of hidden layers and the connections between the neurons. A fully-connected neural network with three input neurons was chosen for the example mathematical description and is depicted in Fig. 2. The neural network comprises one input layer with three neurons, one hidden layer with four neurons and one output layer with two neurons.

The mathematical description of the aforementioned ANN is presented below:

$$O_3 = w_0 + \sum_{i=1}^{1} w_i \times O_{i-1} = w_0 + w_1 \times O_0$$

$$O_4 = w_2 + \sum_{i=3}^{3} w_i \times O_{i-2} = w_2 + w_3 \times O_1$$

$$O_5 = w_4 + \sum_{i=5}^{5} w_i \times O_{i-3} = w_4 + w_5 \times O_2$$

$$O_6 = w_6 + \sum_{i=7}^{9} w_i \times O_{i-4} = w_6 + w_7 \times O_3 + w_8 \times O_4 + w_9 \times O_5$$

$$O_7 = w_{10} + \sum_{i=11}^{13} w_i \times O_{i-8} = w_{10} + w_{11} \times O_3 + w_{12} \times O_4 + w_{13} \times O_5$$

$$O_8 = w_{14} + \sum_{i=15}^{17} w_i \times O_{i-12} = w_{14} + w_{15} \times O_3 + w_{16} \times O_4 + w_{17} \times O_5$$

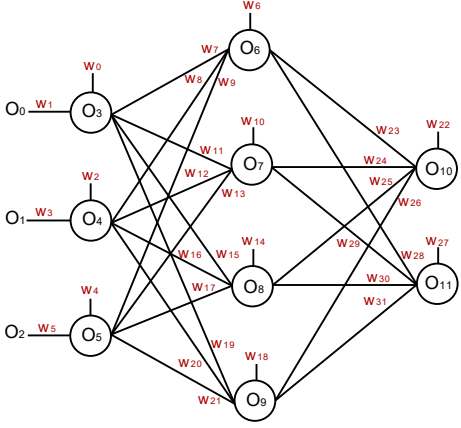

Fig. 2. Example of a fully-connected artificial neural network which served as a basis for the extended mathematical model. The neural network presents: one input layer with three neurons: ($O_3$, $O_4$, $O_5$), one hidden layer with four neurons: ($O_6$, $O_7$, $O_8$, $O_9$) and one output layer with two neurons: ($O_{10}$, $O_{11}$).

$$O_9 = w_{18} + \sum_{i=1}^{21} w_i \times O_{i-16} = w_{18} + w_{19} \times O_3 + w_{20} \times O_4 + w_{21} \times O_5$$

$$O_{10} = w_{22} + \sum_{i=23}^{26} w_i \times O_{i-1} = w_{22} + w_{23} \times O_6 + w_{24} \times O_7 + w_{25} \times O_8 + w_{26} \times O_9$$

$$O_{11} = w_{27} + \sum_{i=2}^{31} w_i \times O_{i-2} = w_{27} + w_{28} \times O_6 + w_{29} \times O_7 + w_{30} \times O_8 + w_{31} \times O_9 \quad (1)$$

This model serves as basis for any feed-forward fully-connected ANN of any dimension. The topology file is uploaded in a 624-bit memory. Following, the neural network's weights and biases are separately stored in a 10.24 kbit dual-port memory in the chip.

In order to understand the ANN's maximum capability, a Matlab model was implemented in floating point and used as a reference. From now on, the Matlab code will be referred to as the golden code, which provides a set of golden outputs. Due to its floating-point representation, it provides much higher precision than a fixed-point representation. The output results of this model are therefore considered to be correct and used as reference. The ANN was also fully described in C and high-level synthesis (HLS) was used to obtain the equivalent register transfer level (RTL) implementation. The ANN's performance in floating point (golden code) and fixed point (C code) are then compared. A conceptual representation of the modelling and comparison procedures is depicted in Fig. 3.

An ANN example was used in order to determine the number of fractional bits needed for the on-chip implementation. The on-chip ANN has to be carefully designed considering availability of resources and area, therefore, the number of fractional bits is an important decision in the ANN design. A large input layer of 10 neurons, 4 hidden layers of 8 neurons each and a 6-neuron output layer ANN was analyzed. A set of random coefficients (weights and biases) sampled from a uniform distribution was used. The ANN's input is given by the 20-bits TDC output codes, which were simulated as random numbers sampled from a uniform distribution across different ranges: [0, 300000], [300000, 600000], [600000, 900000], [900000, 1000000]. The same ANN configuration with the same values was used in the HLS test bench. At the end, the output performance files of both Matlab golden code and HLS were compared as illustrated in Fig. 4. The result indicated a relative rounding error of less than 0.03 % across all TDC input ranges for 8 fractional bit representation.

## III. SYSTEM ARCHITECTURE

Smarty is part of a system-on-chip (SoC) in TSMC's 16 nm FinFET process and it interfaces with a Risc-V processor through an AXI bus. The system comprises a timing block, consisting of 10 TDCs and a processing block consisting of a fully-reconfigurable feed-forward ANN. ANN readout and configuration are also performed through the AXI bus. The SoC PLL provides the clock for the entire system. The ANN was operated at 100 MHz. A block diagram of the main blocks implemented in Smarty is presented in Fig. 5. Three different isolated supply voltage domains are provided for the design: a dedicated power supply for the TDC voltage-controlled ring oscillators (VDD_RING) enabling individual control of the oscillation frequency of the TDCs, a supply voltage dedicated to the ANN (VDD_ANN) allowing the user to test the ANN independently, and a dedicated supply voltage for all the remaining circuits of Smarty (VDD_CORE). Hereafter, the main blocks are described in detail.

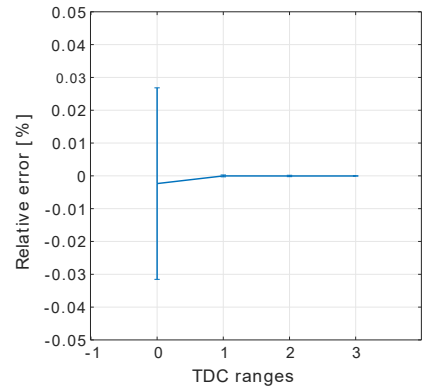

Fig. 4. Relative rounding error of Smarty ANN's outputs. The error was obtained by comparing the golden outputs in floating point and the fixed point outputs. Three different TDC ranges were analyzed: 0 corresponds to [0, 300000], 1 corresponds to [300000, 600000], 2 corresponds to [600000, 900000] and 3 corresponds to [900000, 1000000].

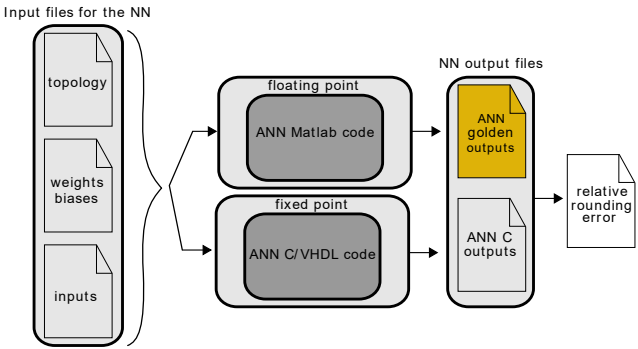

Fig. 3. Conceptual representation of Smarty's ANN modelling and comparison procedure. The ANN's performance in Matlab (floating point) is compared to that of the HLS-inferred system which is fixed point bounded. At the end, the relative error between the two different results is calculated.

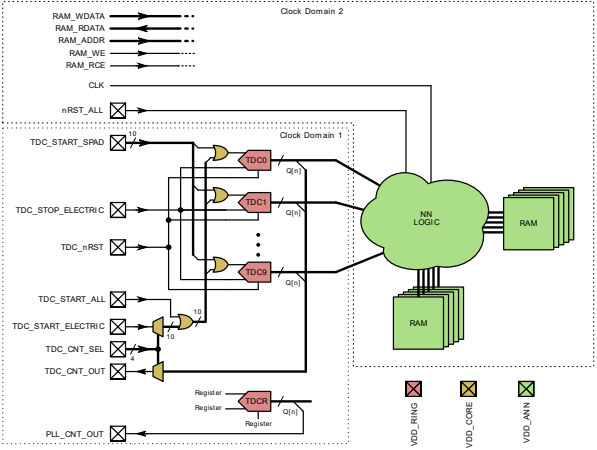

Fig. 5. Smarty block diagram. The chip comprises 10 TDCs whose outputs provide the ANN's inputs. Two dual port memories are used for storing the weights, biases and the ANN outputs. All TDCs can be bypassed and the ANN can be used as a stand-alone structure. Three different isolated supply domains are provided: VDD_RING, VDD_CORE and VDD_ANN.

*A. Time-to-digital converter*

All 10 TDCs present in Smarty are based on the same architecture, a ring topology that comprises a voltage-controlled oscillator (VCO), an asynchronous ripple-counter and a thermometer decoder, as shown in Fig. 6.

The VCO is based on four-delay stages connected in a ring, comprising buffers, inverters and NAND gates as illustrated in Fig. 7. An enable signal (EN) is formed by the rising edge of a START and STOP signal; the VCO starts oscillating when this signal is asserted. The oscillation stops on the EN's falling edge and the frozen state is read using the four outputs Q<0:3>. The 20-bit asynchronous ripple counter keeps track of the number of oscillations through the ring and returns the most-significant bits (MSB), while the least-significant bits are given by the four outputs Q<0:3> through a thermometer decoder. In order to reduce power consumption, the VCO's outputs are buffered and are only available when EN_read is asserted, the only exception being the Q<3> signal, which acts as a clock signal for the counter and has an always-on buffer in order to balance the load along the ring. The final TDC result is given by the following equation:

$$N_{result} = 4 \times N_{coarse} + N_{fine}, \quad (2)$$

where $N_{coar}$ is the counter value and $N_{fine}$ is the decoded fine bit value.

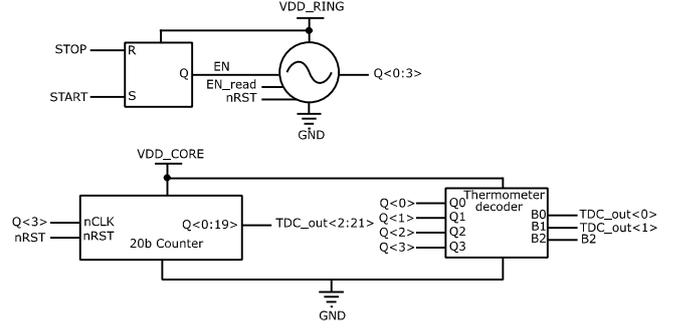

Fig. 6. TDC's main building blocks: a VCO that returns the four phases of the TDC (Q<0:3>), a 20-bit asynchronous ripple counter that returns the TDC's MSBs and a thermometer decoder which returns the decoded values of the TDC's LSBs.

Each TDC can be independently read out. Signals TDC_START_ELECTRIC and TDC_STOP_ELECTRIC are generated by the FPGA during the electrical testing, so as to perform single-shot measurements. TDC_START_ELECTRIC is generated with the same frequency. An adjustable phase with respect to the STOP signal allows different impulse widths to be fed into the TDC, thereby sweeping a larger TDC range. Another option is to start all the TDCs together by using TDC_START_ALL and TDC_STOP_ELECTRIC. In this way, the measurement time is decreased and all TDCs can measure the same impulse width. The TDC's oscillation frequency can be easily determined by monitoring the 7[th] counter bit of each TDC, selected using the TDC_CNT_SEL signal. The TDC's operating principle is illustrated in a timing diagram in Fig. 8.

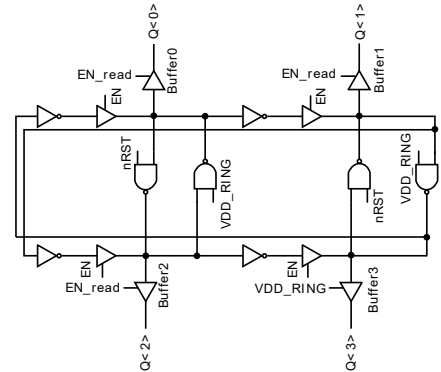

Fig. 7. TDC's ring oscillator structure [20].

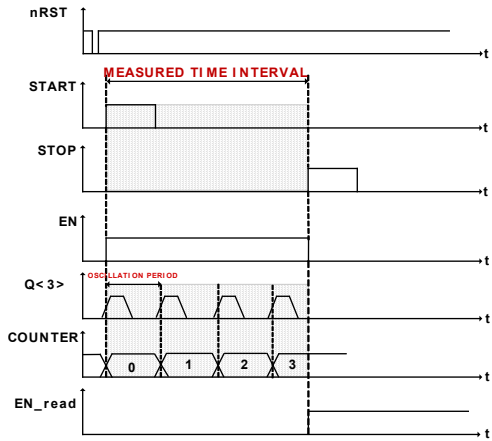

Fig. 8. TDC operating principle. Q<3> signal represents the counter CLK. The VCO oscillates during the period when the EN signal is set to high.

### B. Feed-forward ANN

The ANN comprises three main memory blocks: a 4096 bits memory, which contains the description parameters for all the neurons, a 10 kbit coefficient memory which comprises the values for all the weights and biases and a 624 bit memory for the topology file. The ANN also comprises 4 processors that are fully synthesized through HLS and are used to accelerate the operations performed by the neural network. A control logic unit implements all the necessary sequential steps that are described in the behavioral code and it is fully inferred by the HLS tool. A conceptual diagram of the neural network is depicted in Fig. 9. The on-chip ANN implementation requires the use of many resources and area, therefore, constraints in terms of the maximum number of neurons and coefficients that can be used were imposed. As a result, the ANN can benefit from a maximum of 128 neurons in total, along with 1024 weights and biases in a fully-connected configuration. The ANN's Si area is 89.79 μm × 182.16 μm and the TDC bank is 20 μm × 250 μm, both designed in 16 nm FinFET CMOS technology.

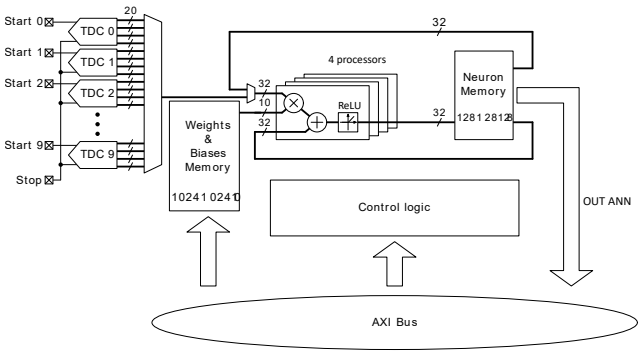

Fig. 9. Conceptual representation of the neural network implementation. The ANN's main blocks are: weights and bias memory, neuron memory and the control logic unit. An AXI bus provides the communication with the ANN.

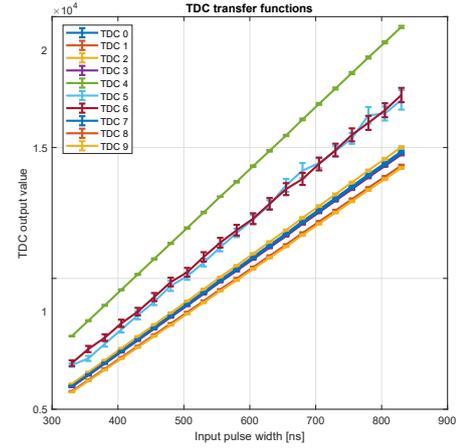

Fig. 10. Transfer functions of all ten TDCs. Measurements performed over a range of 800 ns.

TABLE I
TDCs DNL AND INL VALUES OF AN AVERAGE LSB OF 53.5 PS

| TDC | DNL [LSB] | INL [LSB] |
|---|---|---|
| TDC0 | -0.19/0.15 | -0.77/0.90 |
| TDC1 | -0.11/0.16 | -0.15/1.18 |
| TDC2 | -0.29/0.29 | -1.02/2.17 |
| TDC3 | -0.12/0.13 | -1.13/0.37 |
| TDC4 | -0.45/0.55 | -1.99/0.87 |
| TDC5 | -0.34/0.36 | -1.04/1.66 |
| TDC6 | -0.43/0.39 | -1.42/1.14 |
| TDC7 | -0.14/0.15 | -0.52/0.90 |
| TDC8 | -0.22/0.23 | -0.51/0.46 |
| TDC9 | -0.13/0.12 | -0.63/0.25 |

## IV. PERFORMANCE CHARACTERIZATION RESULTS

The Smarty chip was characterized from different perspectives. Initially, the TDCs were characterized, followed by the ANNs, finally, a coincidence measurement setup was built with two boards placed in front of each other in order to reconstruct the radioactive source position, along the line connecting them.

### A. TDC performance

Electrical tests were performed to determine the transfer functions of each TDC. The transfer function was determined by using the TDC_START_ALL and TDC_STOP_ELECTRIC signal. A large range of impulse widths was covered by sending with an FPGA the TDC_STOP_ELECTRIC and TDC_START_ALL with different phases with respect to each other. The transfer functions of all ten TDCs are shown in Fig. 10. The bin width of each TDC was measured by monitoring the oscillation period of the $7^{th}$ counter bit. The TDCs present an average LSB of 53.5 ps.

The nonlinearities of the TDCs were measured through a code density test, illuminating a photodetector connected to the TDC with white light and accumulating multiple frames. The DNL and INL results of all TDCs are presented in Table I.

## B. Genetic algorithm training

The ANN has been trained in Python 3.6 by using the PyTorch open-source machine learning framework and genetic algorithms [19], [20], [21]. After exploring different training approaches, the use of genetic algorithms proved to be a good solution for the problems that need to be solved by using Smarty's ANN. The training starts with a set of $n$ individuals which are part of a group called population. The number of individuals in each population is chosen by the user and is different from one problem to another. There is no specific number that gives the best result in the end. Initially, the evolution starts from a random population with a certain number of individuals. As in real life, each individual is characterized by genes organized in chromosomes, in this case weights and bias values. A loss function is defined and each individual is evaluated with it during the training process. In the end, the best individual in the final generation is chosen as the final solution.

The mutation and crossover are important parameters which influence the final result. The crossover is the chromosome combination process between two parents whose result is a new offspring. The crossover is usually implemented between the parents which present the best chromosomes so that the new generation has more advanced individuals. A mutation is a genetic operator that takes place after the crossover occurred and represents a random change in one or more sections of the new offspring's chromosome. In this way, there is a high chance that the algorithm converges faster and it is not stuck in local minima. As in the case of the population size and generation number, there is no fixed value for the mutation and crossover parameters which returns the best performance. These values are chosen by the user depending on the problem that is being solved. An illustration of the genetic algorithm main steps used in this framework is depicted in Fig. 11. The GA was used to train a fixed ANN topology across all generations as depicted in Fig. 12. Each individual in each generation represents an ANN. The mutation and crossover operators were used to optimize the weights and biases and minimize the loss function. All the following results are obtained through training with genetic algorithms and their respective chosen parameters will be mentioned along with each presented result.

## C. Measurement setups

To assess the performance of Smarty, three different measurement setups have been created. The first setup, presented in Fig. 13 is a PCB (green), was designed to accommodate the testing of the entire SoC and can be used for electrical measurements only. The second PCB (black) is an interface board which comprises a set of SMAs which allow the interface and control of Smarty's TDCs. In this way, the TDCs can be accessed and triggered externally or interfaced with photodetectors, such as silicon photomultipliers. The last PCB is an application dedicated design which comprises a set of five silicon photomultipliers (Hamamatsu S14160/S14161 series). The A-SiPMs are placed in a line arrangement, one next to each other as depicted in Fig. 14. A dedicated 3D-printed support was designed for this board in order to allow its attachment to an optical table to keep it in a stable position. Each A-SiPM has a corresponding amplifier and comparator whose output is available through SMA connectors.

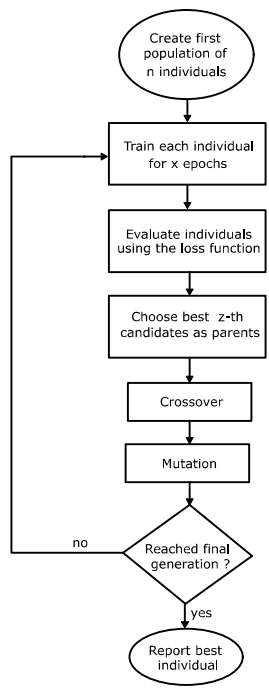

Fig. 11. Main steps of the genetic algorithm used in Smarty's ANN. Adapted from [20].

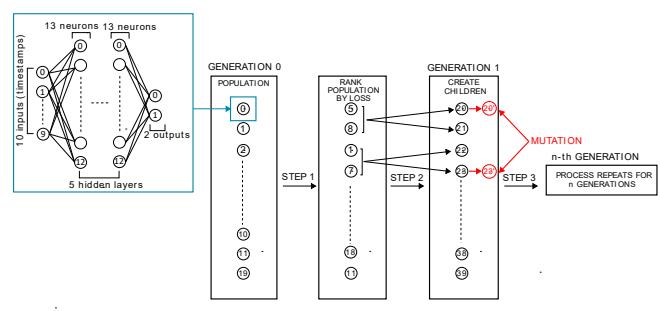

Fig. 12. An ANN example. Each circle in each generation is an ANN of a fixed topology. Each generation has 20 individuals. The biases and weights recombine to create the children of the next generation. A mutation occurs randomly in the chromosomes and it is represented in the red circles.

The sensor board is shown in Fig. 14. In this way, the A-SiPMs signals can be sent as inputs to the TDCs in Smarty. A connection between the interface board and the board which comprises the photodetectors is done through coaxial cables of equal lengths as shown in Fig. 15 b). The A-SiPM features 50% PDE at 450 nm for an excess bias voltage of 2.7 V, corresponding to a total voltage of 40.7 V [19].

### D. Electro-optical ANN performance evaluation

The first measurements performed with Smarty's ANN were single-shot optical measurements. The A-SiPMs were illuminated with a 375 nm picosecond diode laser (PiL037-FC). The board was interfaced with an FPGA and a START signal was generated by the FPGA serving as a trigger for the laser controller. The arrival of the A-SiPM output pulse with respect to a STOP signal, which was also generated by the FPGA, was measured. In a single-shot measurement setup, different delays between the START and STOP signals are measured by the TDCs. A specific TDC code with some variations is generated at each measured interval. The TDC codes corresponding to all five A-SiPMs were read out and a set of 10,000 frames were accumulated for each single-shot measurement. A picture with the measurement setup is presented in Fig. 15. The data of all the single-shot points were transferred to the PC and used to train the ANN using the Python 3.6 training flow with GA. The TDC outputs are connected to the ANN, therefore all the TDC codes were used as input data for the ANN. The chosen topology for the training contains 5 inputs, 5 hidden layers with 13 neurons each and one output neuron. For the GA algorithm training the following parameters were chosen: 10 generations with 30 individuals each (each individual is a neural network of the topology described before), uniform crossover (each gene is chosen for either parent with a certain probability in order to be transferred to one of the two children), 0.2% mutation rate and 10000 epochs. The inputs of the ANN are the TDC codes obtained through the optical single-shot measurement while the output of the ANN is the delay between the START and STOP signals of the TDC (called EN width pulse). The final training results are presented in Fig. 16. The training results show the capability of the ANN to distinguish all 6 different single-shot points solely using the TDC codes. The loss is calculated for each individual in each generation and at the end, the individual with the best average loss across all the frames is chosen. In this case, the loss is calculated as the absolute error of the ANN output. It can clearly be noted that the loss value improved across all 10 generations. At the end of the training, the weights and biases were extracted and transferred to the on-chip ANN. As discussed earlier, the on-chip design makes use of a fixed-point representation, therefore the weights and biases required off-chip quantization. The final on-chip ANN performance is illustrated in Fig. 17. Initially, a naive quantization method was used by directly converting to fixed point through rounding. However, in this case, the results were far from the desired values. A second approach that consisted of scaling the weights and bias before converting to fixed-point representation featured much better results. The performance can be further improved by exploring different quantization methods as well as performing quantized-aware training. The quantized ANN was able to distinguish all 6 single-shot measured points as well as to successfully interpolate three never-before-seen points (28, 33, 47 represented by green dots in Fig. 17).

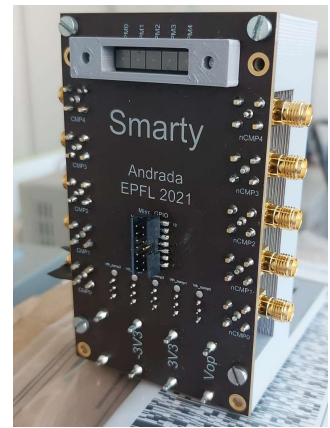

Fig. 14. Sensor board which comprises five 4 mm × 4 mm A-SiPM (Hamamatsu S14160/S14161 series). An amplifier (ADA4807) and a comparator (LT1394) are connected to each A-SiPM; the output of the comparator is made available through SMA connectors. The board can be interfaced with the board of Fig. 13 through coaxial cables as shown in Fig. 15 b).

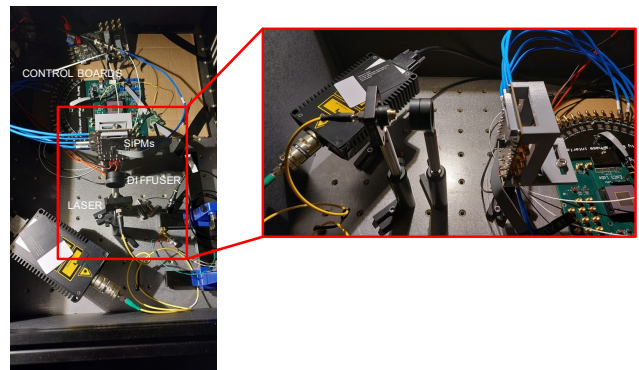

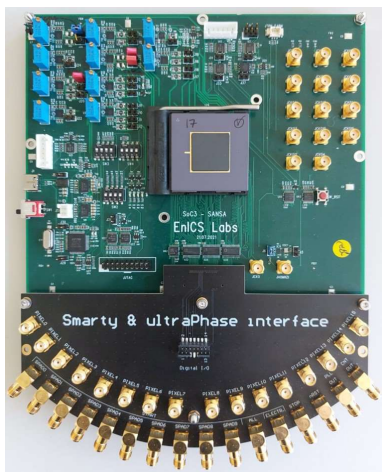

Fig. 13. SoC board (green PCB) connected to the interface board (black PCB). A set of SMA connectors is available for testing and interfacing with other circuits.

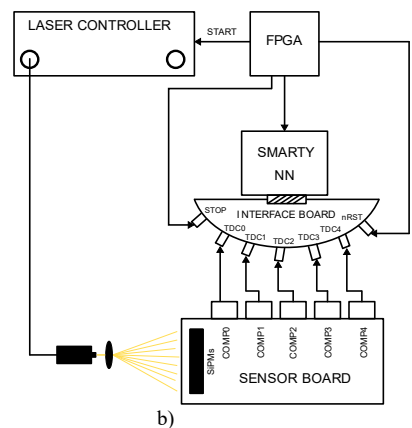

Fig. 15. a) Optical measurement setup for the single-shot measurements. A sensor board which comprises five A-SiPMs is used. All SiPMs are illuminated with a 375 nm picosecond laser. A diffuser (DG20-220-MD) is used between the laser and A-SiPMs in order to assure uniform illumination across all five photodetectors. b) Conceptual representation of the measurement setup presented in a).

## V. ANN PERFORMANCE IN A COINCIDENCE MEASUREMENT SETUP

The ANN designed in Smarty can be used in different measurement setups dedicated to different applications selected by the user. In the scope of this paper, the performance of the ANN was tested in a coincidence measurement setup as well, with applicability for PET scenarios. The goal is to test the ANN's capabilities for the reconstruction of the position of a radioactive source. One of the goals of a PET system is to reconstruct the position of the annihilation point along a line-of-response with very good precision. Through the following simulations and measurements, the ANN's performance capabilities have been tested aiming for a good reconstruction along the X axis.

### A. Simulation setup

Firstly, the performance of the ANN was tested by using synthetic data generated with the Geant4 platform [20]. The simulated model emulates the behavior of the gamma interaction inside a 4 mm × 4 mm × 20 mm LYSO scintillator. This specific scintillator dimension was chosen so that it can match the total area occupied by all five A-SiPMs of the sensor board (sidewise readout). A spherical $^{22}$Na source with a diameter of 3 mm and an intensity of 3.7 MBq placed on disk case with a diameter of 25 mm and a thickness of 6 mm was used in the simulation environment. The gamma rays emitted by the source interact with the scintillators placed in opposition. Upon interaction, the scintillators produce a burst of visible photons. The resulting detection times recorded at the output surface of the scintillator (in this case the 20 mm surface covered by the A-SiPMs) are used as training data for the ANN. Each source position was simulated and the photon arrival times at the output surface of each scintillator recorded individually for each source position. A conceptual block diagram which depicts the simulation environment is shown in Fig. 18.

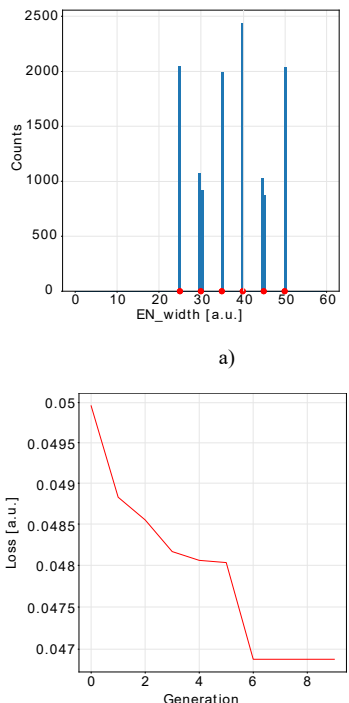

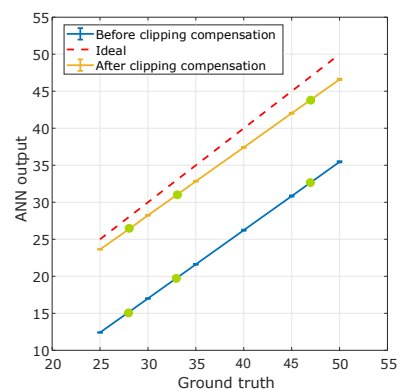

Fig. 16. a) Histogram of the EN width pulse estimation at the output of the neural network when presented with blind validation input frames for 6 different values for the single-shot optical measurement. Ground truth is shown by red dots. b) Average loss of the best performing individual in each generation of the GA. The EN width is presented in arbitrary units and it represents the number of clock cycles chosen during the optical measurement (one clock cycle is 5 ns).

Fig. 17. Smarty ANN on-chip performance in the single-shot measurement setup. The coefficient quantization effects on the ANN's output are depicted in yellow and blue lines. Yellow: naive quantization method which rounds the coefficients. Blue: coefficients are clipped within a desired range. The ANN interpolated three never-before-seen points, represented by green dots.

All the timestamps that reach the two scintillators corresponding to detector1 and detector 2 are used for the ANN training. Each scintillator has an ID which is 1 or 2 so that the spatial information on the arrival of gamma photons is retained. Each source position is simulated one-at-a-time and generates its own dataset containing the spatial coordinates of the source, $X_{source}$ and $Y_{source}$, the time of arrival of the photons at the detection surface of the scintillator and the ID of the scintillation (detector 1 and detector 2). The recorded timestamps were then transformed into TDC codes in order to be consistent with the real measurement scenario in which the ANN receives directly the final codes of all 10 TDCs. Each TDC records a single timestamp per frame per A-SiPM. The length of the frame is set so that multiple TDCs fire. As before, the ANN has been trained in Python 3.6 by using solely timestamps. The organization of the timestamps in preparation for the ANN training is the following:

- all the timestamps are sorted and all their corresponding source position coordinates, $X_{source}$ and $Y_{source}$, are retained;
- the data is organized in exposure frames;
- if all 10 TDCs fired, all ten timestamps are kept for each frame;[1]
- Each A-SiPM has an ID from 1 to 10, each timestamp is assigned to a specific A-SiPM depending on its position at the detection surface of the scintillator;
- At the end, the dataset is organized in a training set which contains 80% of all the data and a validation set which contains 20% of the entire dataset.

As before, the ANN was trained using the GA using solely the TDCs' timestamps. The GA's training framework used for the simulation of the coincidence setup is as follows:

- the first generation contains a population of 20 individuals. Each individual is an ANN with a certain topology that will be further presented along with the simulation results;
- the Adam optimization algorithm [24] is used along with a large number of epochs with a variable learning rate that decreases from 0.01 in the first epoch to 0.001 in the last epoch, whereas it is changed across different training trials;
- the algorithm is ran over 30 generations with a uniform crossover and small mutation rate (< 1%);
- each frame of the dataset contains the TDCs' timing information and their corresponding radioactive source position in the plane;
- finally, the average loss of each individual is calculated for each generation and the individual with the best average loss value from the last generation is reported. The loss function is defined as the distance between the estimated source position and the actual source position in space and it is calculated as follows:

$$loss = \|P_1 P_2\| = \sqrt{(X_1 - X_2)^2 - (Y_1 - Y_2)^2}, \quad (3)$$

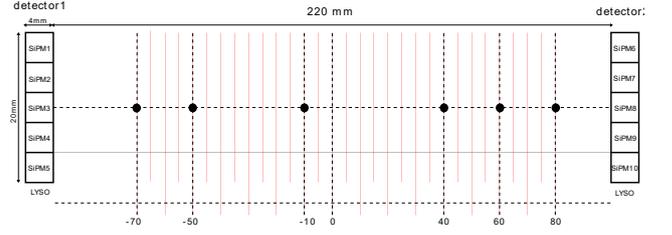

Fig. 18. Geant4 simulation setup with two scintillators placed in coincidence at a distance of 220 mm. Two 4 mm × 4 mm × 20 mm LYSO scintillators were used. The black dots represent the radioactive source positions that are simulated in the Geant4 environment one-at-a-time.

where $P_1$ is the point corresponding to the actual source position in space with its respective coordinates ($X_1$ and $Y_1$), and $P_2$ is the point corresponding to the ANN estimated source position with its respective coordinates ($X_2$ and $Y_2$).

The ANN's reconfigurability consists in changing its topology within the design limitations. In general, there is no best choice in terms of the ANN's topology, crossover or mutation rate. The choice is made by the user and depends on the problem at hand. In this sense, the exploration space is very large and cannot be covered entirely. Therefore, the ANN's performance in a coincidence measurement setup for radioactive source position reconstruction was analyzed by training two different feed-forward topologies: fully-connected narrow-deep and wide-shallow as depicted in Fig. 19. The topology of the feed-forward narrow-deep configuration comprises 10 input neurons, 5 hidden layers of 13 neurons per layer and 2 output neurons, while the feed-forward wide-shallow configuration comprises 10 input neurons, 1 hidden layer with 70 neurons and 2 output neurons. The performance of the best performing topology, the narrow-deep fully-connected ANN with a mutation rate of 0.2% is depicted in Fig. 20. In both cases, the training has been performed with different mutation rates, 1% and 0.2% respectively. The ANN presents better results in the case of a training performed with 0.2%. For both topologies, the ANN was able to distinguish six different radioactive source positions along the X axis. The performance of both ANN topologies is presented in Table II.

TABLE II
ANN PERFORMANCE VALUES

| TOPOLOGY | MUTATION RATE [%] | FINAL LOSS [mm] |
|---|---|---|
| Narrow-deep | 1 | 6.41 |
| Narrow-deep | 0.2 | 4.75 |
| Wide-shallow | 1 | 10.55 |
| Wide-shallow | 0.2 | 9.08 |

---

[1] There are also frames with fewer number of timestamps due to the fact that not all TDCs fired. The first timestamp for each A-SiPM in one frame is kept. A jitter of 120 ps FWHM has been imposed over the synthetic data.

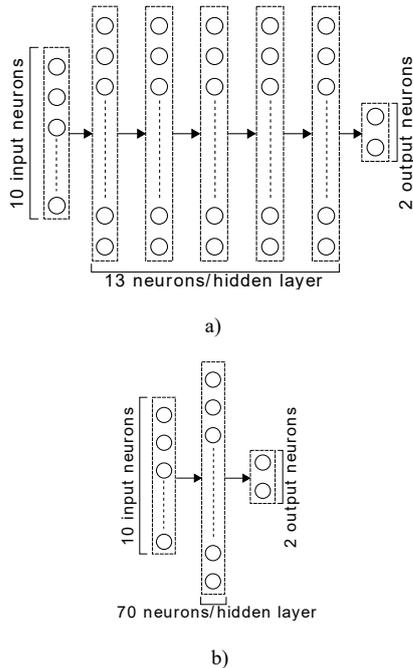

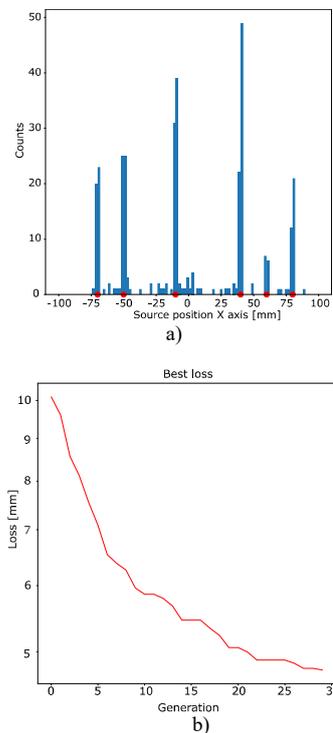

Fig. 19. a) Narrow-deep fully-connected ANN with 10 input neurons, 5 hidden layers of 13 neurons each and 2 output neurons. b) Wide-shallow fully-connected NN with 10 input neurons, 1 hidden layer of 70 neurons and 2 output neurons.

Fig. 20. Narrow-deep feed-forward fully-connected ANN trained with a mutation rate of 0.2%. a) Histogram of the X coordinate estimation at the output of the neural network when presented with never-before-seen validation input frame for 6 radioactive source positions.

### B. On-chip performance

The coincidence measurement setup is presented in Fig. 21. Two A-SiPM sensor boards are placed in coincidence at a distance of 220 mm from each other. A $^{22}$Na radioactive source is placed on an optical rail and its position along the X axis can be changed manually. The 10 comparator outputs from the two boards are connected to the Smarty board via equal length coaxial cables. The control of the boards is performed with the aid of an FPGA. Thousands of frames are accumulated with the TDCs at each source position along the X axis. All the data acquired with the TDCs for different radioactive source positions is used for the training of the ANN. The ANN has been trained using the GA as described before. Compared with the previous training, a classification approach was used due to its better performance in this specific case.

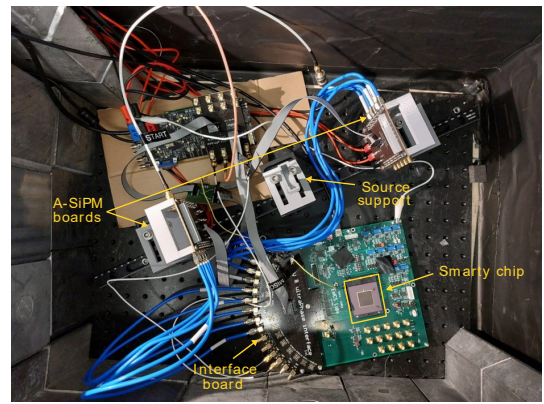

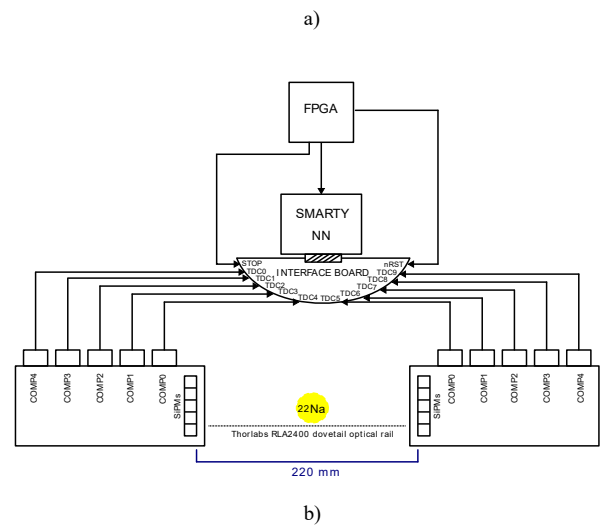

Fig. 21. Smarty coincidence measurement setup. a) Two A-SiPM boards are placed in coincidence at a distance of 220 mm from each other. Each board is coupled with a LYSO scintillator of 4 mm × 4 mm × 20 mm. A $^{22}$Na source can be moved along the X axis between the two sensor boards on a dovetail optical rail. b) Conceptual representation of the coincidence measurement setup.

|  | Predicted position [mm] | | |
|---|---|---|---|
| Classes | -120 | -57 | 65 |
| -120 | 473 | 12 | 8 |
| -57 | 21 | 158 | 77 |
| 65 | 19 | 107 | 112 |

a)

|  | Predicted position [mm] | | |
|---|---|---|---|
| Classes | -120 | -57 | 65 |
| -120 | 471 | 13 | 9 |
| -57 | 23 | 171 | 62 |
| 65 | 23 | 115 | 96 |

b)

Fig. 22. Confusion matrices of the ANN classification results for 2 different source positions placed along the X axis, -57 mm and 65 mm. The non-valid frames are marked with -120. a) Floating point representation, b) quantized model. The numbers represent the number of frames.

|  | Predicted position [mm] | | | |
|---|---|---|---|---|
| Classes | -120 | -57 | 0 | 65 |
| -120 | 477 | 3 | 1 | 2 |
| -57 | 30 | 96 | 64 | 65 |
| 0 | 18 | 66 | 98 | 50 |
| 65 | 7 | 79 | 73 | 93 |

a)

|  | Predicted position [mm] | | | |
|---|---|---|---|---|
| Classes | -120 | -57 | 0 | 65 |
| -120 | 477 | 3 | 3 | 0 |
| -57 | 29 | 90 | 74 | 62 |
| 0 | 10 | 61 | 108 | 53 |
| 65 | 8 | 78 | 75 | 91 |

b)

Fig. 23. Confusion matrices of the ANN classification results for 3 different source positions placed along the X axis, –57 mm, 0 mm and 65 mm. The non-valid frames are marked with -120. a) Floating point representation, b) quantized model. The numbers represent the number of frames.

The ANN's output is divided in a set of classes depending on the number of source positions used for training. The ANN returns the class the source position corresponds to instead of its absolute value in mm. The ANN comprises 10 input neurons, 5 hidden layers with 13 neurons each and 3 output neurons. The performance of the ANN is reported as a confusion matrix. For the training process, frames in which only the TDCs corresponding to one detector fired were considered non-valid frames and a class called -120 mm, which is a position beyond the distance between the two photodetectors, was associated to them. Non-valid frames are frames in which only one of the detectors (detector 1 or detector 2) fired, therefore, no coincidence can be performed by the ANN.

The classification results of two radioactive source positions placed at –57 mm and 65 mm along the X axis between the two sensor boards are presented in Fig. 22 for both floating point and quantized models. The performance is quantified using two parameters: accuracy and precision.

$$ACCURACY = \frac{TP + TN}{TP + TN + FP + FN},$$
$$PRECISION = \frac{TP}{TP + FP}$$
(4)

where TP are true positives, TN true negatives, FP false positives and FN false negatives. The average accuracy and precision of the floating-point representation is 83.59% and 68.69% respectively, while the quantized model presents an accuracy of 83.48% and a precision of 68.59%. From the results presented in Fig. 22, it can be observed that the ANN was able to clearly distinguish between valid and non-valid frames with high degree of certainty. Both positions, –57 mm and 65 mm have been distinguished by the ANN, however, it favors the former, most likely due to unequal number of frames in the input dataset and bias differences between the two detectors that led to a slight increase in the count rate on the SiPMs on the left. This effect is more evident in the quantized model.

The same analysis has been repeated for three different source positions, –120 mm (represented by non-valid frames), -57 mm, 0 mm and 65 mm. The ANN's classification results are shown in Fig. 23. In this case, the floating-point model has an accuracy of 81.26% and a precision of 53.70%, while the quantized model has an accuracy of 81.34% and a precision of 53.88%. As in the previous case, the ANN distinguished with a very high degree of certainty the difference between valid and non-valid frames. The two models have a similar performance.

*C. Hardware performance evaluation*

The execution time of the ANN was measured considering an ANN with 10 input neurons, 5 hidden layers with 13 neurons per hidden layer and 3 output neurons. For a 105 MHz clock, the ANN has an execution time of 22.44 μs. Considering a total number of 1710 operations, the ANN executes 76.15 MOPS. However, the ANN was designed to run at a maximum frequency of 500 MHz which results in a maximum performance of 363 MOPS. At a frequency of 100 MHz, which proved to be sufficient for the current experiment, the ANN itself consumes 0.4 mW, which is equivalent to 190 GOPS/W. The entire SoC highlighting the Smarty design is shown in Fig. 24.

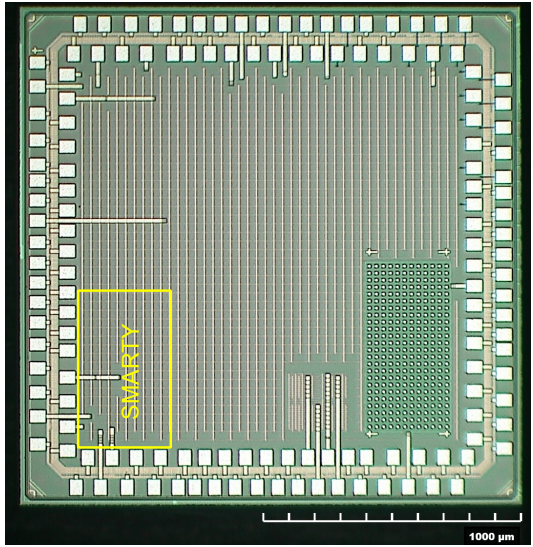

Fig. 24. Overall SoC micrograph. Approximate representation of Smarty location is shown in the yellow rectangle.

## VI. Conclusions

Smarty is a fully-integrated and fully-reconfigurable feed-forward artificial neural network with ten TDCs designed in a 16 nm FinFET CMOS technology, intended to operate in a PET coincidence measurement setup with reduced system complexity and data throughput. The system exhibits a high degree of flexibility, the user being able to decide the ANN topology in accordance to the problem at hand, within a set of constraints. The average LSB of the TDCs is 53.5 ps, while the ANN operates at 363 MOPS with an efficiency of 190 GOPS/W. Smarty was validated in an experimental coincidence setup where it successfully distinguished between six different positions of a radioactive source when using a floating-point implementation and three different source positions with a quantized model. In all the test cases, the ANN was able to filter with high degree of certainty frames which were considered as being non-valid, i.e. frames without timestamps in which no coincidence occurred.

Smarty represents the first step towards reducing output data throughput and overall system complexity by bringing a level of preprocessing close to the photodetector.

Conflicts of Interest: The authors declare no conflict of interest.

## References

[1] S. Bickley, H. Chan and B. Torgle, "Artificial intelligence in the field of economics," *Scientometrics,* vol. 127, pp. 2055-2084, 2022.

[2] J. G. Greener, S. M. Kandathil, L. Moffat and D. T. Jones, "A guide to machine larning for biologists," *Nature Reviews Molecular Cell Biology,* vol. 23, pp. 40-55, 2022.

[3] S. J. Russell and P. Norvig, Artificial Intelligence: A Modern Approach, Prentice Hall, 2009.

[4] E. S. Soegoto, R. D. Utami and Y. A. Hermawan, "Influence of artificial intelligence in automotive industry," *Journal of Physics,* vol. 1402, no. 6, 2019.

[5] D. Simon, Evolutionary Optimization Algorithms, Biologically-Inspired and Population-Based Approaches to Computer Intelligence, John Wiley and Sons Inc., 2013.

[6] E. Altay Varol and B. Alatas, "Performance analysis of multi-objective aritificial intelligence optimization algorithms in numerical association rule mining," *J Ambient Intell Human Computation,* pp. 3449-3469, 2020.

[7] Y. LeCun, Y. Bengio and G. Hinton, "Deep learning," *Nature,* vol. 521, pp. 436-444, 2015.

[8] C. Floyd, "An artificial neural network for SPECT image reconstruction," *IEEE Transactions on Medical Imaging,* vol. 10, no. 3, pp. 485-487, 1991.

[9] M. T. Munley, C. E. Floyd, J. E. Bowsher and R. E. Coleman, "An artificial neural network approach to quantitative single photon emission computed tomographic reconstruction with collimator, attenuation, and scattered compensation," *Medical physics,* vol. 21, no. 12, 1994.

[10] K. Gong, E. Berg, S. R. Cherry and J. Qi, "Machine Learning in PET: From Photon Detection to Quantitative Image Reconstruction," *Proceedings of the IEEE,* vol. 108, no. 1, pp. 51-68, 2020.

[11] I. Häggström, C. R. Schmidtlein, G. Campanella and T. J. Fuchs, "DeepPET: A deep encoder-decoder network for directly solving the PET image reconstruction inverse problem," *Medical Image Analysis,* vol. 54, pp. 253-262, 2019.

[12] K. Gong, J. Guan and C. Q. J. Liu, "PET Image Denoising Using a Deep Neural Network Through Fine Tuning," *IEEE Transactions on Radiation and Plasma Medical Sciences,* vol. 3, no. 2, p. 153.161, 2019.

[13] S. I. Kwon, R. Ota and E. Berg, "Ultrafast timing enables reconstruction-free positron emission imaging," *Nature Photonics,* vol. 15, pp. 914-918, 2021.

[14] E. Venialgo, S. Mandai, T. Gong, D. R. Schaart and E. Charbon, "Time estimation with multichannel digital silicon photomultipliers," *Physics in medicine and biology,* vol. 60, pp. 2435-2452, 2015.

[15] P. Carra, M. G. Bisogni, E. Ciarrocchi and et. al., "A neural network-based algorithm for simultaneous event positioning and timestamping in monolithic scintillators," *Physics in Medicine and Biology,* vol. 67, no. 13, 2022.

[16] M. A. Miller, *Vereos Digital PET/CT Performance,* Philips, 2016.

[17] B. A. Spencer, E. Berg, J. P. Schmall, N. Omidvari and et. al., "Performance Evaluation of the uEXPLORER Total-Body PET/CT Scanner Based on NEMA NU 2-2018 with Additional Tests to Characterize PET Scanners with a Long Axial Field of View," *Journal of Nuclear Medicine,* vol. 62, no. 6, pp. 861-870, 2021.

[18] G. A. Prenosil, H. Sari, M. Fuerstner and et. al., "Performance Characteristics of the Biograph Vision Quadra PET/CT system with long axial field of view using the NEMA NU 2-2018 Standard," *Journal of Nuclear Medicine,* vol. 63, no. 3, 2021.

[19] K. Man, K. Tang and S. Kwong, "Genetic algorithms: concepts and applications in engineering design," *IEEE Transactions on Industrial Electronics,* vol. 43, no. 5, pp. 519-534, 1996.

[20] K. Guo, M. Yang and H. Zhu, "Application research of improved genetic algorithm based on machine learning in production scheduling," *Neural Computing and Applications,* vol. 32, 2020.

[21] A. Hassanat, K. Almohammadi, E. Alkafaween, E. Abunawas, A. Hammouri and V. B. S. Prasath, "Choosing Mutation and Crossover Ratios for Genetic Algorithms - A Review with a New Dynamic Approach, " *Information,* vol. 10, no. 2, 112, 2019.

[22] Hamamatsu, *Low breakdown voltage type MPPC for scintillation detector, S14160/S14161,* Hamamatsu, 2020.

[23] *Geant4,* https://geant4.web.cern.ch/.

[24] P. K. Diederik and B. Jimmy, Adam: A Method for Stochastic Optimization, 2015.

[25] A. A. AbdulHamed, M. A. Tawfeek and A. E. Keshk, "A genetic algorithm for service flow management with budget constraint in heterogeneous computing," *Future Computing and Informatics Journal,* vol. 3, no. 2, pp. 341-347, 2018.

[26] C. Veerappan, J. Richardson, R. Walker and et. al., "A 160 x 128 single-photon image sensor with on-pixel 55ps 10b time-to-digital converter," *IEEE International Solid-State Circuits Conference,* pp. 312-314, 2011.